\documentclass[aps,prd,reprint,floatfix,superscriptaddress]{revtex4-1}
\usepackage{graphicx}
\usepackage{color}
\usepackage{cancel}
\usepackage{amsmath}
\usepackage{amsfonts}
\usepackage{amssymb}
\newcommand{\be}{\begin{equation}}
\newcommand{\ee}{\end{equation}}

\begin{document}
\title{Relaxation to quantum equilibrium for Dirac fermions \mbox{in the de Broglie-Bohm pilot-wave theory}}
\author{Samuel Colin}
\email{s.colin@griffith.edu.au}
\affiliation{Centre for Quantum Dynamics, Griffith University, Brisbane, QLD 4111, Australia}
\affiliation{Perimeter Institute for Theoretical Physics, 31 Caroline Street North, ON N2L2Y5, Waterloo, Canada}
\begin{abstract}
Numerical simulations indicate that the Born rule does not need to be postulated in the de Broglie-Bohm pilot-wave theory, but arises dynamically (relaxation to quantum equilibrium).
These simulations were done for a particle in a two-dimensional box whose wave-function 
obeys the non-relativistic Schr\"odinger equation and is therefore scalar. 
The chaotic nature of the de Broglie-Bohm trajectories, thanks to the nodes of the wave-function which act as vortices, 
is crucial for a fast relaxation to quantum equilibrium.
For spinors, we typically do not expect any node. However, in the case of the Dirac equation, the de Broglie-Bohm velocity field has vorticity even in the absence of nodes. 
This observation raises the question of the origin of relaxation to quantum equilibrium for fermions.
In this article, we provide numerical evidence to show that Dirac particles also undergo relaxation, by simulating the evolution of various non-equilibrium 
distributions for two-dimensional systems (the 2D Dirac oscillator and the Dirac particle in a spherical 2D box). 
\end{abstract}
\maketitle
\section{Introduction}
In the de Broglie-Bohm pilot-wave theory \cite{debroglie28,bohm521,bohm522}, each element of an ensemble (consisting of a single particle) is described by a wave-function $\psi(t,\vec{x})$ ($R(t,\vec{x})e^{iS(t,\vec{x})/\hbar}$ in the polar representation) but also by a position $\vec{x}(t)$. 
The wave-function always evolves according to the Schr\"odinger equation
\be\label{nrse}
i\hbar\partial_t\psi(t,\vec{x})=-\hbar^2 m^{-2}\Delta\psi(t,\vec{x})+V(\vec{x})\psi(t,\vec{x})~,
\ee
whereas the evolution of $\vec{x}(t)$ is determined by the standard guidance equation
\be\label{guidance}
\vec{v}(t)=m^{-1}\vec{\nabla}S(t,\vec{x})\big|_{\vec{x}=\vec{x}(t)}~.
\ee
The de Broglie-Bohm theory reproduces the predictions of standard quantum mechanics provided that the particle positions are distributed according to
\be\label{equi}
\rho(t,\vec{x})=|\psi(t,\vec{x})|^2 
\ee
over the ensemble. De Broglie's dynamics (defined by Eq. (\ref{nrse}) and Eq. (\ref{guidance}), by contrast to Bohm's dynamics \cite{costva}) 
ensures that this condition will hold at all times, if it holds for some initial time $t_0$, a property which is referred to as quantum equilibrium \cite{valentini91a,valentini91b} 
or equivariance \cite{durr92}. 
There are (at least) four different attitudes towards this condition (and some of them are not mutually exclusive): 
Valentini argues that it arises dynamically \cite{valentini91a,valentini91b,valentini-phd}, 
D\"urr, Goldstein and Zangh\`{\i} argue that we happen to live in a typical universe \cite{durr92}, Bricmont takes 
it as a third postulate \cite{bricmont}, while Wiseman suggests that it can be derived as a Bayesian prior from a principle of indifference \cite{wiseman2007}. 
In this article, we will be concerned with the first attitude, which opens a door to a possible new physics, that of quantum non-equilibrium (see for instance \cite{valentini08} and \cite{cowi}).

If this view is considered seriously, one has to explain why we don't see quantum non-equilibrium today. 
This is done by invoking the idea of relaxation to quantum equilibrium: if some non-equilibrium distributions existed in the past, they are quickly 
driven dynamically to quantum equilibrium. 
Various numerical simulations for two-dimensional systems (\cite{valentini05,cost10,toruva}) have shown that non-equilibrium distributions rapidly relax to quantum equilibrium on a coarse-grained level, provided that the wave-function has enough complexity. By complexity, we mean that two trajectories originating from neighborhood points 
should quickly diverge (which is referred to as chaos in a loose sense). The existence of nodes has been first connected to chaos in \cite{frisk97}. 
So when we say that the wave-function needs to be complex enough, we mean that one needs to superpose a few energy modes in order to get a few nodes.
Nodes are also associated to vorticity: the circulation of the velocity field around a closed curved can only be non-zero if there is a node inside the curve.
More recently, a better understanding of chaos, through `nodal-point--X-point complex' (where the X-point is an unstable point associated to the node),  
has also been gained \cite{efthymiopoulos}.

The relaxation simulations have been performed for non-relativistic scalar particles. No simulation has ever been done for fermions. 
And there is a good reason to do simulations for fermions, because we typically don't expect any node for spinors, no matter how many energy modes are being superposed. 
The reason is the following. For scalar particles, we have nodes where $\mathfrak{Re}(\psi)=0$ and $\mathfrak{Im}(\psi)=0$. In 2D, nodes are located at the intersection 
of the curves defined by the two previous equations and we typically have a few nodal points (in 3D, we would have nodal lines). 
For a positive-energy spin-$\frac{1}{2}$ fermion, described by a Pauli spinor $\begin{pmatrix}\psi_1\\\psi_2\end{pmatrix}$, nodes are located at the intersection 
of four curves (or surfaces): $\mathfrak{Re}(\psi_1)=0$, $\mathfrak{Re}(\psi_2)=0$, $\mathfrak{Im}(\psi_1)=0$ and $\mathfrak{Im}(\psi_2)=0$. 
Therefore we typically don't have any node in two or three dimensions. 
This raises the question of relaxation to quantum equilibrium. At first, it seems problematic for spinors. 

But the velocity field for spinors (for Dirac particles for instance) is essentially different from the standard velocity-field for non-relativistic scalar particles: 
it has vorticity even in the absence of nodes. If vorticity is the crucial ingredient for relaxation to quantum equilibrium, we expect relaxation to quantum equilibrium 
for Dirac particles too. Actually a previous work by Colin and Struyve \cite{cost10} gives some support to that idea. 
There the evolution of non-equilibrium distribution in a 2D square box is simulated, the wave-function being nodeless (except at the boundaries). 
Relaxation is very poor in that case, except if  `artificial vortices' are added, such that the alternative (and non-standard) velocity field exhibits vorticity even in the absence of nodes.
This is always possible and is referred to as the non-uniqueness problem of the de Broglie-Bohm theory \cite{deotto98,wiseman2007}. 
This example can be a toy-model for spinors; however one can argue with reason that these non-standard velocity fields are unnatural 
and that this example lacks features which are inherent to spinors (for instance the spinorial density can't vanish at the boundaries).

That is the main question that we will address in this paper. In order to do that, we will consider 2 two-dimensional systems for `confined' Dirac particles: the first one is the so-called 
Dirac oscillator, the second one is a Dirac particle confined in a spherical box. In each case, we study a superposition of positive-energy solutions.
We will show that the first system, in which 8 modes are superposed, exhibit fast relaxation, while the second system, in which 6 modes are superposed, also undergoes 
relaxation to quantum equilibrium (although it is a slower one). 
In the second system, we also give an example of a non-trivial spinor for which there will never be complete relaxation.
 
This article is organized in the following way. 
Section II contains the basic details about the Dirac equation in $3+1$ and $2+1$ dimensional spacetimes, as well as the corresponding pilot-wave theory.
In section III and IV we study relaxation to quantum equilibrium for Dirac spinors. 
We conclude in Section V.

We use units in which $\hbar=c=1$.
\section{Pilot-wave theory for a Dirac fermion}
In this section, we recall the essential properties of the Dirac equation in spacetimes of dimension $4$ and $3$, and the corresponding pilot-wave theory.
\subsection{$3+1$-dimensional spacetime}
The Dirac equation is 
\be
i\frac{\partial}{\partial t}\psi(t,\vec{x})=-i\vec{\alpha}\cdot\vec{\nabla}\psi(t,\vec{x})+m\beta\psi(t,\vec{x})~,
\ee
where the matrices $\alpha_j$ and $\beta$ must be Hermitian and must satisfy the relations $\{\alpha_j,\alpha_k\}=2\delta_{jk}$, $\beta^2=1$ and $\{\alpha_j,\beta\}=0$.

The Dirac equation can be rewritten in a covariant form by introducing the matrices $\gamma^0=\beta$ and $\gamma^j=\beta\alpha_j$:
\be
(i\gamma^\mu\partial_\mu-m)\psi(t,\vec{x})=0~.
\ee
The $\gamma$-matrices satisfy the relations $\{\gamma^\mu,\gamma^\nu\}=2\eta^{\mu\nu}$, where $\eta^{\mu\nu}=\textrm{diag}(1,-1,-1,-1)$ (Clifford algebra). 
The Pauli-Dirac representation 
\be
\tilde\gamma^0=\begin{pmatrix}1 & 0\\ 0 & -1\end{pmatrix}~, \tilde\gamma^i=\begin{pmatrix}0 & \sigma_i\\ \sigma_i & 0\end{pmatrix}~,
\ee
and the Weyl (or chiral) representation
\be
\gamma^0=\begin{pmatrix}0 & 1\\ 1 & 0\end{pmatrix}~,\gamma^i=\begin{pmatrix}0 & \sigma_i\\ -\sigma_i & 0\end{pmatrix}
\ee
are two common choices of representation for the $\gamma$-matrices. In this article we use the Weyl representation. 
The conserved 4-current is given by 
\be
j^\mu=\bar\psi\gamma^\mu\psi~,
\ee
where $\bar\psi=\psi^\dagger\gamma^0$.

The positive and negative-energy plane-wave solutions are denoted by $u(\vec{p})e^{-iE(\vec{p})t+i\vec{p}\cdot\vec{x}}$ and 
$v(\vec{p})e^{iE(\vec{p})t+i\vec{p}\cdot\vec{x}}$, where $E(\vec{p})=\sqrt{|\vec{p}|^2+m^2}$. 
In the Weyl representation, if one introduces the right-handed and left-handed eigenstates of helicity 
$\chi_R(\vec{p})$ and $\chi_L(\vec{p})$, the plane-wave solutions can be given by 
\begin{eqnarray}
u_R(\vec{p})=\begin{pmatrix}\sqrt{\frac{E-p}{2E}}\chi_R(\vec{p})\\ \sqrt{\frac{E+p}{2E}}\chi_R(\vec{p})\end{pmatrix}~,\label{ur}\\
u_L(\vec{p})=\begin{pmatrix}\sqrt{\frac{E+p}{2E}}\chi_L(\vec{p})\\ \sqrt{\frac{E-p}{2E}}\chi_L(\vec{p})\end{pmatrix}~,\label{ul}\\
v_L(\vec{p})=\begin{pmatrix}\sqrt{\frac{E-p}{2E}}\chi_L(\vec{p})\\ \sqrt{\frac{E+p}{2E}}\chi_L(\vec{p})\end{pmatrix}~,\label{vl}\\
v_R(\vec{p})=\begin{pmatrix}\sqrt{\frac{E+p}{2E}}\chi_R(\vec{p})\\ \sqrt{\frac{E-p}{2E}}\chi_R(\vec{p})\end{pmatrix}~.\label{vr}
\end{eqnarray}
There is an easy way to remember this. We write the Dirac spinor as $\psi=\begin{pmatrix}\psi_L\\\psi_R\end{pmatrix}$. In the limit $m=0$, 
the Dirac equation reduces to a pair of Weyl equations for $\psi_R$ and $\psi_L$. The Weyl equation for $\psi_R$ admits right-handed 
positive-energy solutions and left-handed negative-energy solutions, whereas the Weyl equation for $\psi_L$ admits left-handed 
positive-energy solutions and right-handed negative-energy solutions. That is indeed what we recover from the previous equations 
in the limit $E\rightarrow p$.

In the corresponding pilot-wave theory (first proposed by Bohm in a reply to Takabayashi \cite{bohm53}), 
the particle is described by a 4-spinor $\psi(t,\vec{x})$ together with its position $\vec{x}(t)$. 
The position of the Dirac particle evolves according to the guidance equation
\be
\vec{v}(t)=\frac{\vec{j}(t,\vec{x})}{j^0(t,\vec{x})}\bigg|_{\vec{x}=\vec{x}(t)}=\frac{\psi^\dagger(t,\vec{x})\vec{\alpha}\psi(t,\vec{x})}{\psi^\dagger(t,\vec{x})\psi(t,\vec{x})}\bigg|_{\vec{x}=\vec{x}(t)}~.
\ee
Again the choice of the guidance equation ensures that the relation 
\be
\rho(t,\vec{x})=\sum_{a=1}^{a=4}|\psi_a(t,\vec{x})|^2
\ee
will hold for any time $t$ provided that it holds for some initial time.

Any positive-energy plane-wave solution of momentum $\vec{p}$ , whether right-handed or left-handed, moves with velocity $\vec{p}/E$ whereas negative-energy plane-wave 
solutions move in the opposite direction of momentum.

\begin{figure}
\includegraphics[width=0.4\textwidth]{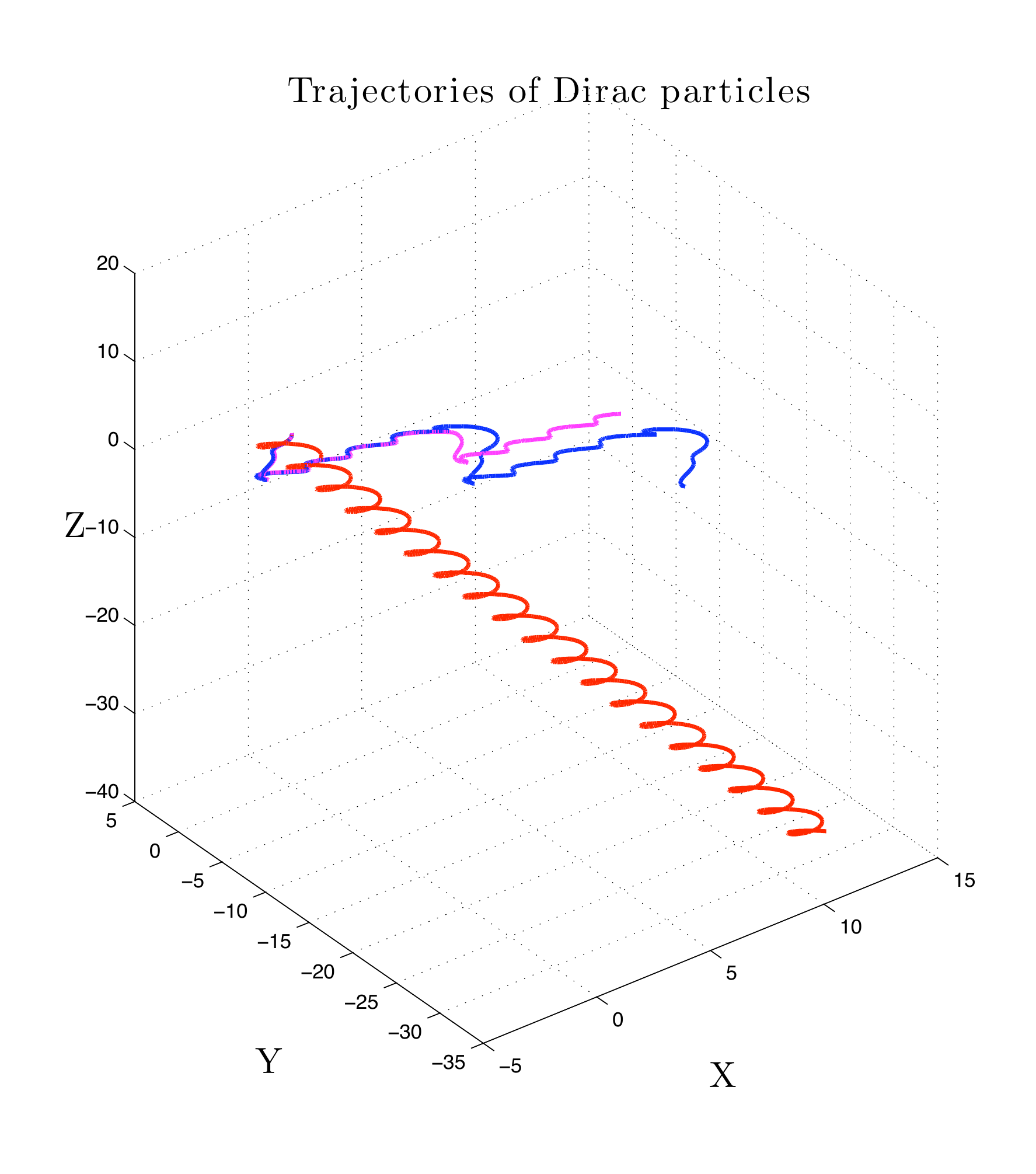}
\caption{\label{fig1}Trajectories of Dirac particles in a $3+1$-dimensional spacetime.
The guiding spinors are obtained from the one given in Eq. (\ref{spinor1}) by using different values of the mass ($3$, $6$ and $9$). 
The particles all start from the origin and $t\in[0,200].$}
\end{figure}

Fig. \ref{fig1} shows the kind of trajectories predicted by the theory. The guiding Dirac spinor in this case is a superposition of three positive-energy 
right-handed plane-wave solutions
\begin{eqnarray}\label{spinor1}
\psi(t,\vec{x})=\frac{1}{\sqrt{3}}(u_R(\vec{p}_1)e^{-iE_1t+i\vec{p}_1\cdot\vec{x}}+e^{i4}u_R(\vec{p}_2)e^{-iE_2t+i\vec{p}_2\cdot\vec{x}}\nonumber\\
+e^{i9}u_R(\vec{p}_3)e^{-iE_3t+i\vec{p}_3\cdot\vec{x}})~,\quad\quad
\end{eqnarray}
where
\be 
\vec{p}_1=(1,0,1),~\vec{p}_2=(-1,-2,-1)\textrm{ and }\vec{p}_3=(1,-1,1)
\ee
and $u_R$ is defined at Eq. (\ref{ur}).

\subsection{$2+1$-dimensional spacetime}
In a $2+1$-dimensional spacetime, the Dirac equation is
\be
i\frac{\partial}{\partial t}\psi(t,x,y)=(-i\alpha_1\frac{\partial}{\partial x}-i\alpha_2\frac{\partial}{\partial y}+m\beta)\psi(t,x,y)~.
\ee
We can take $\alpha_1=\sigma_1$, $\alpha_2=\sigma_2$ and $\beta=\sigma_3$. 

The covariant form is 
\be
(i\gamma^0\partial_t+i\gamma^1\partial_x+i\gamma^2\partial_y-m)\psi(t,x,y)=0~,
\ee
where the $\gamma$-matrices are defined by
\be
\gamma^0=\sigma_3,~\gamma^1=\sigma_3\sigma_1,~\gamma^2=\sigma_3\sigma_2~.
\ee

For the plane-wave solutions, we have 
\be
u(p_x,p_y)=\sqrt{\frac{E+m}{2E}}\begin{pmatrix}1\\\frac{p_x+ip_y}{E+m}\end{pmatrix}e^{-iEt+ip_x x+ip_y y}~,
\ee
and 
\be
v(p_x,p_y)=\sqrt{\frac{E+m}{2E}}\begin{pmatrix}\frac{-p_x+ip_y}{E+m}\\1\end{pmatrix}e^{iEt+ip_x x+ip_y y}~,
\ee
which are respectively positive and negative-energy solutions. 

\begin{figure}
\includegraphics[width=0.5\textwidth]{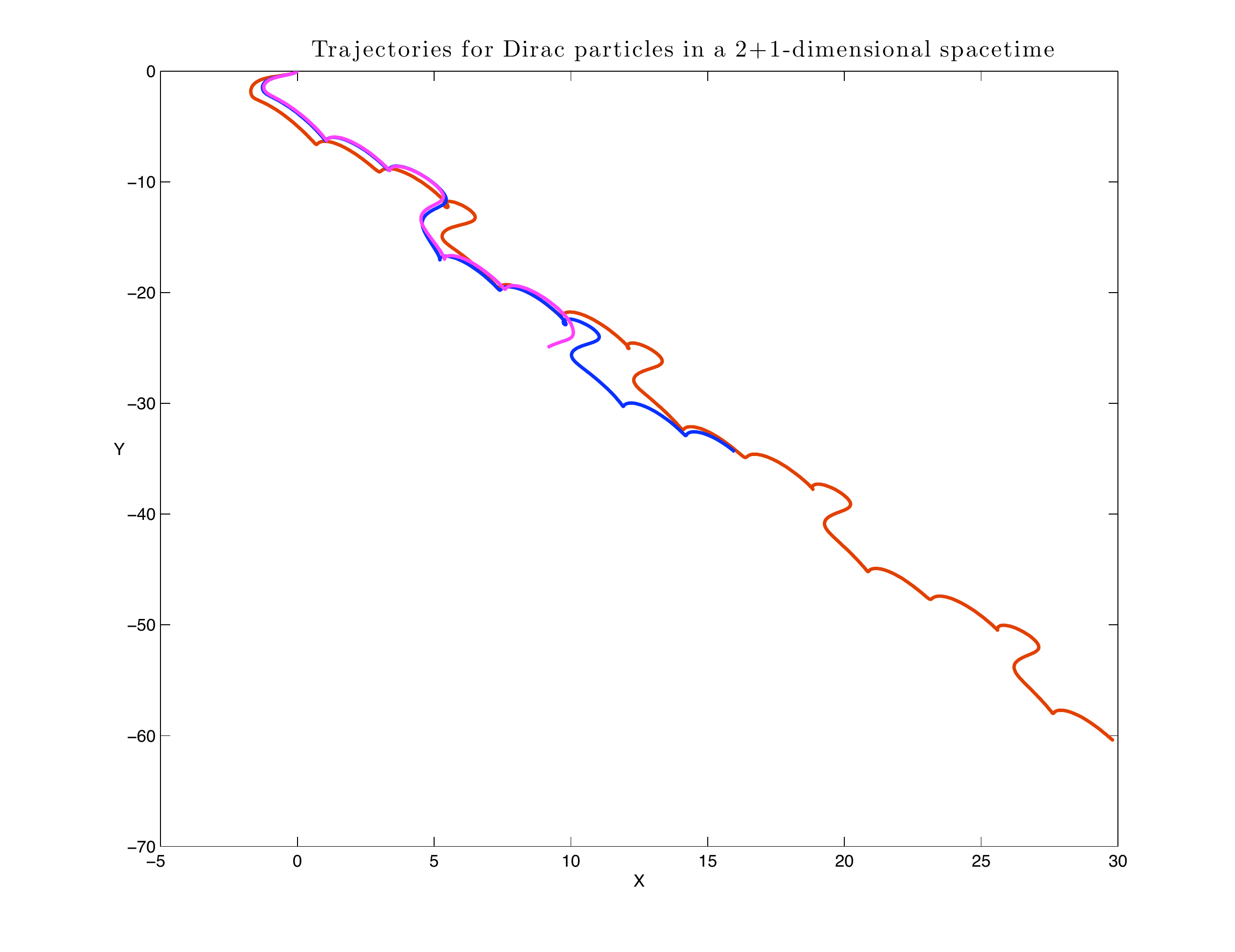}
\caption{\label{fig1b}Trajectories of Dirac particles in a $2+1$-dimensional spacetime.
The guiding spinors are obtained from the Dirac spinor given in Eq. (\ref{spinor2}) by using different values of the mass ($3$, $6$ and $9$). 
The particles all start from the origin and $t\in[0,200].$}
\end{figure}

In Fig. \ref{fig1b}, we plot a few trajectories of particles guided by the spinor
\begin{eqnarray}\label{spinor2}
\psi(t,\vec{x})=\frac{1}{\sqrt{3}}(u(\vec{q}_1)e^{-iE_1t+i\vec{q}_1\cdot\vec{x}}+e^{i4}u(\vec{q}_2)e^{-iE_2t+i\vec{q}_2\cdot\vec{x}}\nonumber\\
+e^{i9}u(\vec{q}_3)e^{-iE_3t+i\vec{q}_3\cdot\vec{x}})~,\quad\quad
\end{eqnarray}
where
\be 
\vec{q}_1=(1,0),~\vec{q}_2=(-1,-2)\textrm{ and }\vec{q}_3=(1,-1)~,
\ee
for different values of the mass.

The circulation of the velocity field 
\be
\oint\vec{v}\cdot\vec{dl}
\ee
can be non-zero, indicating vorticity, even in the absence of nodes. 
Indeed the non-relativistic limit of the Dirac equation gives a correction (spin-term) to the standard de Broglie-Bohm velocity field \cite{bohm93}.
\section{Relaxation to quantum equilibrium for the 2D Dirac oscillator}
The equation for the Dirac oscillator is obtained by substituting $\vec{p}$ by $\vec{p}-im\omega\beta\vec{r}$ in the free Dirac Hamiltonian \cite{moshinsky}:
\be
H=\vec{\alpha}\cdot(\vec{p}-im\omega\beta\vec{r})+m\beta~.
\ee
This system reduces to a standard oscillator plus a strong spin-orbit coupling term in the non-relativistic limit 
and some of its generalizations \cite{diracosc3d1,diracosc3d2} are relevant to the study of quarks confinement.

The 2D case, with Hamiltonian
\begin{eqnarray}\label{ham}
H=\begin{pmatrix}0 & p_x-ip_y\\p_x+ip_y&0\end{pmatrix}+im\omega\begin{pmatrix}0&x-iy\\-x-iy&0\end{pmatrix}\nonumber\\
+\begin{pmatrix}m&0\\0&-m\end{pmatrix}~\qquad
\end{eqnarray}
was treated by Villalba in \cite{villalba}. 
We have verified Villalba's construction and we arrive at the same energy eigenstates 
expect that a few intermediate equations are different, which is why we include the derivation here. Also, it will be useful for the case of the Dirac particle in a 2D spherical box.
\subsection{Energy eigenstates}
We first multiply Eq. (\ref{ham}) by $i$ and we note that
\begin{eqnarray}
\partial_x+i\partial_y=(\partial_x\theta+i\partial_y\theta)\partial_\theta+(\partial_x r+i\partial_y r)\partial_r\nonumber\\
=r^{-1}(-\sin{\theta}+i\cos{\theta})\partial_\theta+e^{i\theta}\partial_r\nonumber\\
=ir^{-1}e^{i\theta}\partial_\theta+e^{i\theta}\partial_r~,\\
\partial_x-i\partial_y=(\partial_x\theta-i\partial_y\theta)\partial_\theta+(\partial_x r-i\partial_y r)\partial_r\nonumber\\
=r^{-1}(-\sin{\theta}-i\cos{\theta})\partial_\theta+e^{-i\theta}\partial_r\nonumber\\
=-ir^{-1}e^{-i\theta}\partial_\theta+e^{-i\theta}\partial_r~.
\end{eqnarray}
Therefore we have that:
\begin{eqnarray}\label{eq1}
iH=\begin{pmatrix}0 & -ir^{-1}e^{-i\theta}\partial_\theta+e^{-i\theta}\partial_r\\ir^{-1}e^{i\theta}\partial_\theta+e^{i\theta}\partial_r&0\end{pmatrix}\nonumber\\
-m\omega\begin{pmatrix}0&r e^{-i\theta}\\-r e^{i\theta}&0\end{pmatrix}+i\begin{pmatrix}m&0\\0&-m\end{pmatrix}~.\qquad
\end{eqnarray}
The previous operator is applied on $\displaystyle\begin{pmatrix}\psi_1\\\psi_2\end{pmatrix}$. If we introduce
\be\label{eq1b}
\psi_1=\frac{1}{\sqrt{r}}e^{-i\frac{\theta}{2}}\psi'_1\textrm{ and }\psi_2=\frac{1}{\sqrt{r}}e^{i\frac{\theta}{2}}\psi'_2~,
\ee
we have that
\begin{eqnarray}
(ir^{-1}e^{i\theta}\partial_\theta+e^{i\theta}\partial_r)\psi_1=\nonumber\\
e^{i\frac{\theta}{2}}(ir^{-\frac{3}{2}}\partial_\theta\psi'_1+\frac{1}{\sqrt{r}}\partial_r\psi'_1)~,\label{eq2}\\
(-ir^{-1}e^{-i\theta}\partial_\theta+e^{-i\theta}\partial_r)\psi_2=\nonumber\\
e^{-i\frac{\theta}{2}}(-ir^{-\frac{3}{2}}\partial_\theta\psi'_2+\frac{1}{\sqrt{r}}\partial_r\psi'_2)~.\label{eq3}
\end{eqnarray}
Now we look for energy eigenstates: we make the replacement
\be\label{eq4}
\psi'(t,r,\theta)\rightarrow\psi'(r)e^{-iEt}e^{i k\theta}~,
\ee
where $k$ is half-integer. From Eqs (\ref{eq1}--\ref{eq4}), we find that
\begin{eqnarray}
i(E-m)\psi'_1=(\partial_r+\frac{k_\theta}{r}-m\omega r)\psi'_2~,\nonumber\\
i(E+m)\psi'_2=(\partial_r-\frac{k_\theta}{r}+m\omega r)\psi'_1~,
\end{eqnarray}
from which the following equation for $\psi'_1$ can be deduced:
\begin{eqnarray}\label{diffequ1}
(E^2-m^2)\psi'_1=\nonumber\\-(\partial^2_r+\frac{k}{r^2}(1-k)+m\omega(1+2k)-m^2\omega^2r^2)\psi'_1~.\qquad
\end{eqnarray}
Now we perform the following change of variables
\be
\xi=m\omega r^2~,
\ee
which implies that
\be
\partial^2_r\psi'_1=\partial_r(2m\omega r\partial_\xi\psi'_1)=
(2m\omega\partial_\xi+4m^2\omega^2 r^2\partial^2_\xi)\psi'_1~.
\ee
Eq. (\ref{diffequ1}) becomes
\begin{eqnarray}\label{diffequ2}
\Delta^2\psi'_1=\nonumber\\-(2\partial_\xi+4\xi\partial^2_\xi+\frac{k}{r^2}(1-k)+(1+2k)-\xi)\psi'_1~,\qquad
\end{eqnarray}
where
\be
\Delta^2=\frac{E^2-m^2}{m\omega}~.
\ee
If we start from the ansatz
\be
\psi'_1=exp(-\xi/2)\xi^\alpha \mathcal{P}(\xi)~,
\ee
Eq. (\ref{diffequ2}) reduces to an equation for generalized Laguerre polynomials ($\mathcal{P}(\xi)\rightarrow\mathcal{P}^\mu_n(\xi)$)
\be
\mathcal{P}''+(1+\mu-\xi)\mathcal{P}'+n\mathcal{P}=0~,
\ee 
provided that $\alpha=k/2$ (or $\alpha=1/2-k/2$), $\mu=k-1/2$ (or $-k+1/2$) and
\be
\Delta^2=4n~(\text{or } \Delta^2=4n-4k+2)~.
\ee
The choice of $\alpha$ (which in turn fixes $\mu$ and $\Delta^2$) is fixed by the sign of $k$, 
because the wave-function needs to be regular at the origin.

We can always choose units in which $\hbar=1$, $c=1$ and $m=1$. The frequency $\omega$ remains a free parameter. 
The following table contains the eigenstates of lowest energy for the case $\omega=1$. 
\begin{eqnarray}\label{psi11}
\begin{tabular}{| l | l |}
\hline
$n=1$ & $\psi_1=\frac{e^{-i\sqrt{5}t}e^{-r^2/2}2(1-r^2)}{\sqrt{2\pi}\sqrt{5-\sqrt{5}}}$\\ 
$k=1/2$ & $\psi_2=\frac{4ie^{i\theta}re^{-i\sqrt{5}t}e^{-r^2/2}}{\sqrt{2\pi}\sqrt{5-\sqrt{5}}(1+\sqrt{5})}$\\ 
\hline
$n=1$ & $\psi_1=\frac{e^{-i\theta}e^{-i3t}e^{-r^2/2}r(2-r^2)}{\sqrt{3\pi}}$\\ 
$k=-1/2$ & $\psi_2=\frac{e^{-i3t}e^{-r^2/2}i(r^2-1)}{\sqrt{3\pi}}$\\ 
\hline
$n=2$ & $\psi_1=\frac{e^{-i3t}e^{-r^2/2}(r^4-4r^2+2)}{\sqrt{6\pi}}$\\ 
$k=1/2$ & $\psi_2=\frac{e^{i\theta}e^{-i3t}e^{-r^2/2}ir(-r^2+2)}{\sqrt{6\pi}}$\\ 
\hline
$n=2$ & $\psi_1=\frac{e^{-i\theta}e^{-i\sqrt{13}t}e^{-r^2/2}r(r^4-6r^2+6)}{\sqrt{2\pi}\sqrt{13-\sqrt{13}}}$\\ 
$k=-1/2$ & $\psi_2=\frac{e^{-i\sqrt{13}t}e^{-r^2/2}6i(-r^4+4r^2-2)}{\sqrt{2\pi}\sqrt{13-\sqrt{13}}(1+\sqrt{13})}$\\ 
\hline
\end{tabular}
\end{eqnarray}
\begin{eqnarray}\label{psi12}
\begin{tabular}{| l | l |}
\hline
$n=1$ & $\psi_1=\frac{e^{i\theta}e^{-i\sqrt{5}t}e^{-r^2/2}r(2-r^2)}{\sqrt{\pi}\sqrt{5-\sqrt{5}}}$\\ 
$k=3/2$ & $\psi_2=\frac{2e^{2i\theta}e^{-i\sqrt{5}t}e^{-r^2/2} i r^2}{\sqrt{\pi}\sqrt{5-\sqrt{5}}(1+\sqrt{5})}$\\ 
\hline
$n=1$ & $\psi_1=\frac{e^{-2i\theta}e^{-i\sqrt{13}t}e^{-r^2/2}r^2(-r^2+3)}{\sqrt{\pi}\sqrt{13-\sqrt{13}}}$\\ 
$k=-3/2$ & $\psi_2=\frac{e^{-i\theta}e^{-i\sqrt{13}t}e^{-r^2/2}6ir(r^2-2)}{\sqrt{\pi}\sqrt{13-\sqrt{13}}(1+\sqrt{13})}$\\ 
\hline
$n=2$ & $\psi_1=\frac{e^{i\theta}e^{-i3t}r(r^4-6r^2+6)}{3\sqrt{2\pi}}$\\ 
$k=3/2$ & $\psi_2=\frac{e^{2i\theta}e^{-i3t}ir^2(3-r^2)}{3\sqrt{2\pi}}$\\ 
\hline
$n=2$ & $\psi_1=\frac{e^{-2i\theta}e^{-i\sqrt{17}t}r^2(r^4-8r^2+12)}{\sqrt{6\pi}\sqrt{17-\sqrt{17}}}$\\ 
$k=-3/2$ & $\psi_2=\frac{e^{-i\theta}e^{-i\sqrt{17}t}8ir(-r^4+6r^2-6)}{\sqrt{6\pi}\sqrt{17-\sqrt{17}}(1+\sqrt{17})}$\\ 
\hline
\end{tabular}
\end{eqnarray}
\subsection{Simulations}
For the wave-function, we consider a superposition of the $8$ previous energy eigenstates with equal weights and 
phases given in the following table:
\begin{eqnarray}\label{psi13}
\begin{tabular}{| l | l |}
\hline
$(n,k)$ & Phase\\
\hline
$(1,1/2)$ & $e^{i4.869}$\\
$(1,-1/2)$ & $e^{i1.049}$ \\
$(2,1/2)$ & $e^{i4.291}$\\
$(2,-1/2)$ & $e^{i3.066}$ \\ 
$(1,3/2)$ & $e^{i0.188}$\\
$(1,-3/2)$ & $e^{i1.288}$ \\
$(2,3/2)$ & $e^{i0.219}$\\
$(2,-3/2)$ & $e^{i4.706}$\\
\hline
\end{tabular}
\end{eqnarray}
We consider $5$ different non-equilibrium distributions at time $t=0$. The first one, denoted by $\rho_0(t=0,r,\theta)$, is defined by 
\be\label{rho0}
\rho_0(t=0,r,\theta)=\frac{2\pi\cos^2(\frac{\pi r}{2R_0})}{R_0^2(\pi^2-4)}\quad\textrm{with~}R_0=4~,
\ee
if $r\leq R_0$, and zero otherwise.
The four remaining distributions ($\rho_1,\rho_2,\rho_3$ and $\rho_4$) are built from $\rho_0$, 
except that they are respectively centered at $\vec{x}_1=(2,0)$, $\vec{x}_2=(0,2)$, $\vec{x}_3=(-2,0)$ and $\vec{x}_4=(0,-2)$ and that  $R_0$ is substituted by $R=2$. 
Explicitly we have that: 
\be\label{rhoj}
\rho_j(t=0,\vec{x})=\frac{2\pi\cos^2(\frac{\pi d_i(\vec{x})}{2R})}{R^2(\pi^2-4)}\quad\textrm{with~}R=2~,
\ee
if $d_i(\vec{x})=\sqrt{(x_j(1)-x(1))^2+(x_j(2)-x(2))^2}\leq R$, and zero otherwise.

The evolution of the non-equilibrium distribution can be computed using a method proposed by Valentini and Westman \cite{valentini05} and also 
used later in \cite{cost10} and \cite{toruva}. This method relies on the fact that the following ratio
\be\label{liou}
\displaystyle\frac{\rho(t,\vec{x})}{\psi^\dagger(t,\vec{x})\psi(t,\vec{x})}
\ee
is preserved along a trajectory. 
Therefore, if we want compute $\rho(t,\vec{x})$, we first `backtrack' the initial position 
(that is, we find the initial condition $(t_0,\vec{x}_0)$ which gives $(t,\vec{x})$ as final position). Then we use Eq. (\ref{liou}) in order to find that
\be
\rho(t,\vec{x})=\psi^\dagger(t,\vec{x})\psi(t,\vec{x})\frac{\rho(t_0,\vec{x}_0)}{\psi^\dagger(t_0,\vec{x}_0)\psi(t_0,\vec{x}_0)}~.
\ee
For this simulation, we have used a lattice of $2048\times 2048$ points, uniformly distributed inside the box $[-5,5]\times[-5,5]$, of coordinates 
\begin{eqnarray}
x&=-5+10/4096+10k/2048\nonumber\\
y&=-5+10/4096+10l/2048~,
\end{eqnarray}
with $k,l\in\{0,1,2,\ldots,2047\}$). Note that all the lattice points are inside the box (none of them lies on the boundary of the box), and that is 
why $10/4098$ appears in the expression (and not $10/2048$). 
Each lattice point (representing the final position of a particle) is backtracked to its initial position 
by solving the differential equation 
\be
\frac{d\vec{x}(t)}{dt}=\frac{\psi^\dagger(t,\vec{x})\vec{\alpha}\psi(t,\vec{x})}{\psi^\dagger(t,\vec{x})\psi(t,\vec{x})}\bigg|_{\vec{x}=\vec{x}(t)}
\ee
numerically using the Runge-Kutta-Fehlberg method \cite{press92}. We have imposed a precision on $0.001$ on the backtracked positions. 
Lattice points that give an insufficient precision, or can't be backtracked before a given maximum number of iterations (for instance $100000$), 
are discarded. Hereafter we refer to these lattice points as bad ones (or good ones if the converse is true).

The next step is the coarse-graining, which is done by dividing the box $[-5,5]\times[-5,5]$ in a certain number of non-overlapping coarse-graining cells, 
and averaging the densities corresponding to the lattice-points contained in each coarse-graining cell.
If we divide the box in $32\times32$ non-overlapping coarse-graining cells, the mean percentage of good lattice points 
per coarse-graining cell is $98.47\%$ and the worst cell has $85.06\%$.

For the plots, we have actually used a smooth coarse-graining of the densities (the same that was used in \cite{cost10}) which is similar 
to a standard coarse-graining except that it is defined over overlapping CG cells. 
In this case, all the overlapping CG cells can be obtained from a square cell of side $10/16$ located in the lower left corner 
by translating it by $m\frac{10}{16\times 8}$ in the up-direction, and by $n\frac{10}{16\times 8}$ in the right-direction, where $m,n\in\{0,120\}$. 
The smooth coarse-graining is denoted by 
\be\label{smooth}
\widetilde{\rho}(t,\vec{x})\textrm{ or }\widetilde{(\psi^\dagger\psi)}(t,\vec{x})~
\ee
where $\vec{x}$ is in fact the center of an overlapping CG cell.

\begin{figure}  
\includegraphics[width=0.5\textwidth]{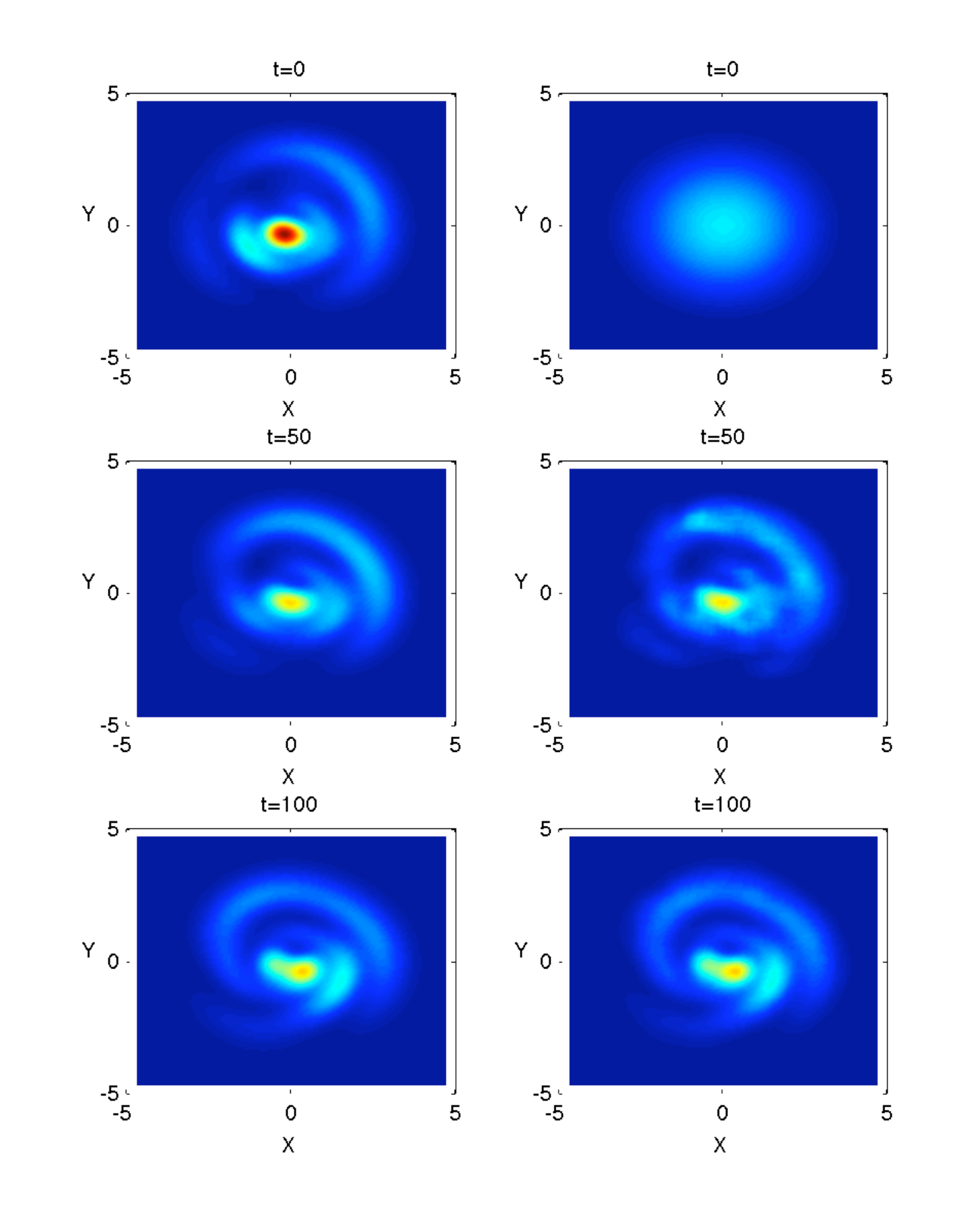}
\caption{\label{fig4a}Density plots for $t\in\{0,50,100\}$. The first column displays the evolution of the smoothly coarse-grained standard density 
($\widetilde{\psi^\dagger\psi}$), while the second column displays the evolution of the smoothly coarse-grained non-equilibrium density ($\widetilde{\rho_0}$). 
The smooth coarse-graining is defined in the paragraph associated to Eq. (\ref{smooth}). The Dirac spinor is defined in Eqs (\ref{psi11},\ref{psi12},\ref{psi13}).}
\end{figure}

\begin{figure}
\includegraphics[width=0.5\textwidth]{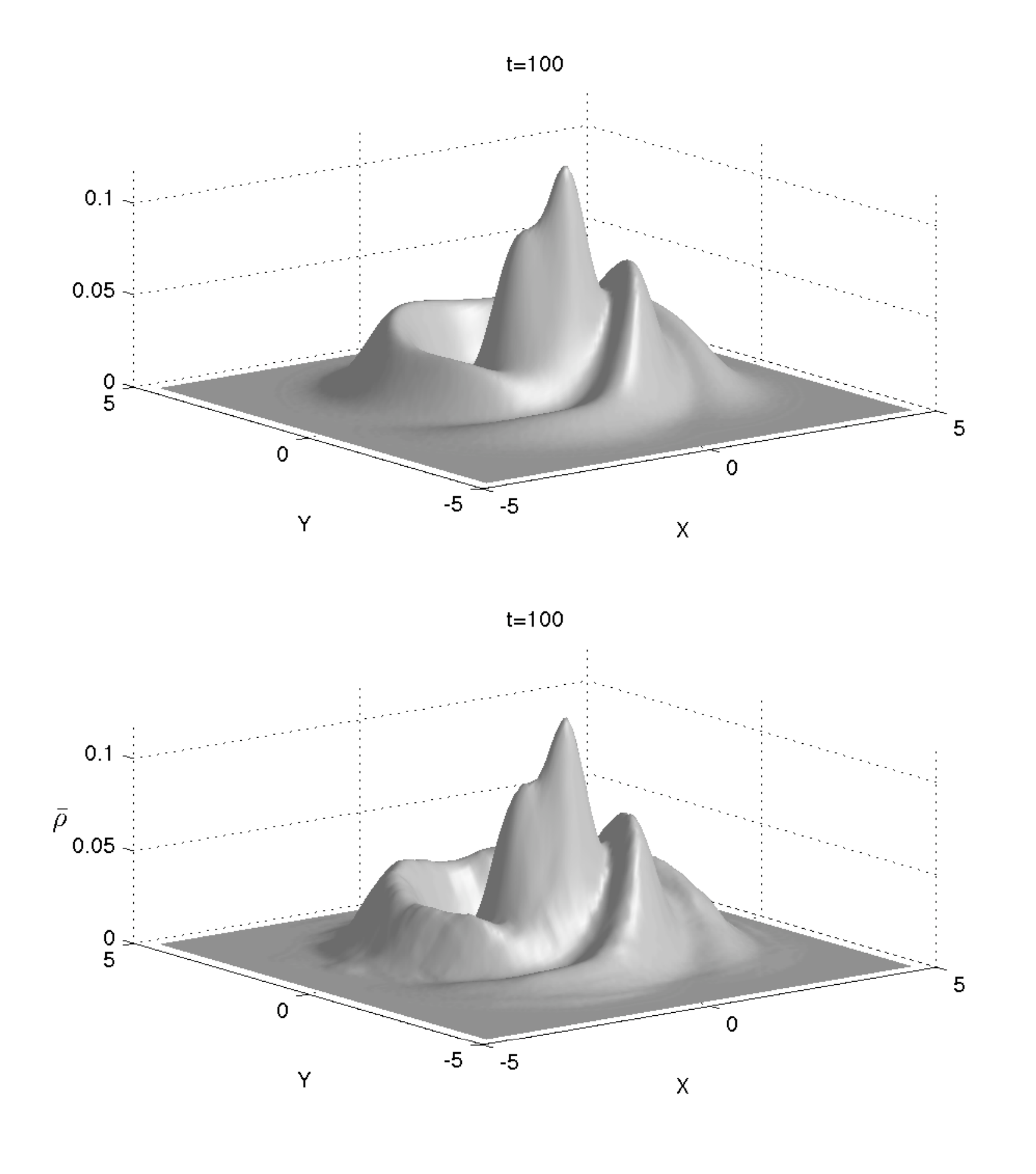}
\caption{\label{fig4b}Surface plot for $\widetilde{\psi^\dagger\psi}$ and $\widetilde{\rho_0}$ at the final time $t=100$. See Eq. (\ref{smooth}) for the definition 
of the smooth coarse-graining and Eqs (\ref{psi11},\ref{psi12},\ref{psi13}) for the definition of the Dirac spinor.}
\end{figure}

\begin{figure}  
\includegraphics[width=0.5\textwidth]{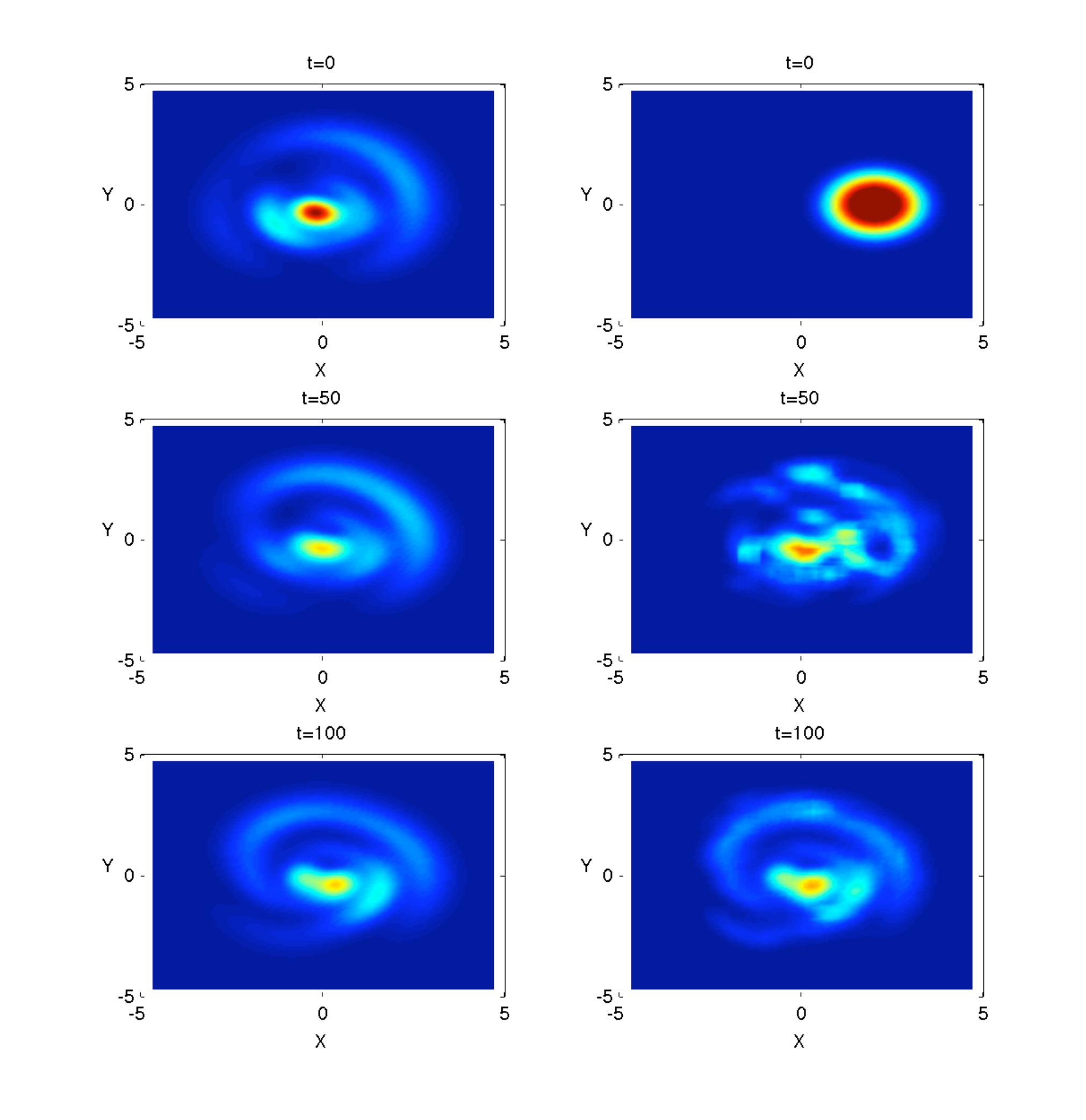}
\caption{\label{fig5a}Density plots for $t\in\{0,50,100\}$. The first column displays the evolution of the smoothly coarse-grained standard density 
($\widetilde{\psi^\dagger\psi}$), while the second column displays the evolution of the smoothly coarse-grained non-equilibrium density ($\widetilde{\rho_1}$). 
The smooth coarse-graining is defined in the paragraph associated to Eq. (\ref{smooth}). The Dirac spinor is defined in Eqs (\ref{psi11},\ref{psi12},\ref{psi13}).}
\end{figure}

\begin{figure}
\includegraphics[width=0.5\textwidth]{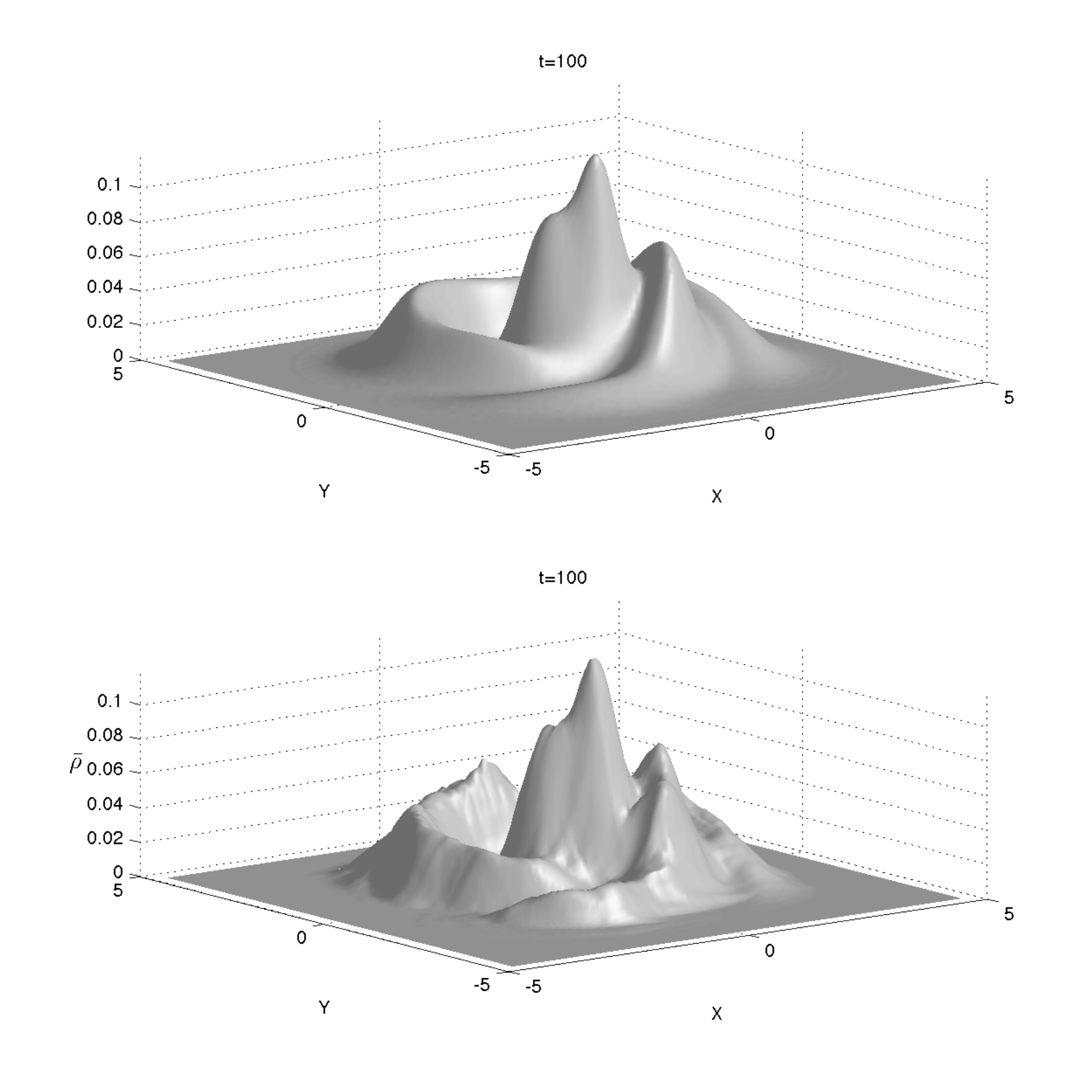}
\caption{\label{fig5b}Surface plot for $\widetilde{\psi^\dagger\psi}$ and $\widetilde{\rho_1}$ at the final time $t=100$. See Eq. (\ref{smooth}) for the definition 
of the smooth coarse-graining and Eqs (\ref{psi11},\ref{psi12},\ref{psi13}) for the definition of the Dirac spinor.}
\end{figure}

The evolution of the non-equilibrium distribution $\rho_0$ is illustrated by a density plot (see Fig. \ref{fig4a}) and also by surface plot for the final time $t=100$ in Fig. \ref{fig4b}. 
We also have the data for two further intermediate steps ($t=25$ and $t=75$). 
Fig. \ref{fig5a} and Fig. \ref{fig5b} illustrate, in the same manner, the evolution of $\rho_1$.
\subsection{Discussion}
For that superposition of eight modes with positive energy, the relaxation to quantum equilibrium is rather fast. In these units, it takes about $\Delta t=100$ 
for the non-equilibrium distribution to resemble $|\psi|^2$, which is about one hundred times the Compton wavelength of the particle divided by the speed of light.
In standard units, it is about $10^{-20}\textrm{s}$ if we take a Compton wavelength of $10^{-14}\textrm{m}$.
\section{Relaxation to quantum equilibrium for a Dirac particle in a 2D spherical box}
We consider a Dirac particle confined to a spherical box of radius $R'$. Inside the spherical box, the potential is negative and constant, outside it is zero:
\be
V(r)=\begin{cases} -|V_0|\quad r\leq R'\\0\quad r>R'\end{cases}~.
\ee
The corresponding 3D case is analyzed in \cite{greiner}.
\subsection{Energy eigenstates}
It is similar to the previous case: we can start from Eq. (\ref{diffequ1}), replace $E$ by $E+|V|$ (where $|V|$ is equal to $|V_0|$ or equal to zero, 
depending on whether we consider the interior solution or the exterior one) and set $\omega=0$. Then we have that
\be
-((E+|V|)^2-m^2)\psi'_1=\partial^2_r\psi'_1+k(1-k)\psi'_1/r^2~,
\ee
where k is half-integer (we define $n=k-1/2$).

We distinguish two cases:
\paragraph{$\kappa^2=(E+|V|)^2-m^2>0$.} 
In that case, the equation to solve is
\be
r^2\partial^2_r\psi'_1+r^2\kappa^2\psi'_1+k(1-k)\psi'_1=0~.
\ee
If we introduce $s=r\kappa$ and $\psi'_1=\sqrt{s}\Phi_1$, we find the following equation:
\be
s^2\partial^2_s\Phi_1+s\partial_s\Phi_1+s^2\Phi_1-n^2\Phi_1=0~,
\ee
which admits Bessel functions of the first and second kind as solutions ($J_{k-\frac{1}{2}}(s)$ and $Y_{k-\frac{1}{2}}(s)$). 

If we put all the pieces together, we have that
\be
\psi'_1=\sqrt{s}(\alpha J_{k-\frac{1}{2}}(s)+\beta Y_{k-\frac{1}{2}}(s))
\ee
and
\be\label{int1}
\psi_1=\sqrt{\kappa}e^{-iEt}e^{i(k-\frac{1}{2})\theta}(\alpha J_{k-\frac{1}{2}}(s)+\beta Y_{k-\frac{1}{2}}(s))
\ee

For $\psi'_2$, we have that
\begin{eqnarray}
\psi'_2=\frac{-i\kappa}{E+|V|+m}(\partial_s-\frac{k}{s})\psi'_1\nonumber\\
=\frac{-i\kappa}{E+|V|+m}(\partial_s-\frac{k}{s})\sqrt{s}\Phi_1\nonumber\\
=\frac{-i\kappa}{E+|V|+m}(\frac{\psi'_1}{2s}-\frac{k}{s}\psi'_1+\sqrt{s}\partial_s\Phi_1)\nonumber\\
=\frac{-i\kappa}{E+|V|+m}(\frac{\psi'_1}{2s}-\frac{k}{s}\psi'_1-\frac{1}{s}(k-\frac{1}{2})\psi'_1+\nonumber\\
\sqrt{x}(\alpha J_{k-\frac{3}{2}}(s)+\beta Y_{k-\frac{3}{2}}(s)))\nonumber~.
\end{eqnarray}
For $\psi_2$:
\begin{eqnarray}\label{int2}
\psi_2=-i\kappa^{\frac{3}{2}}\frac{e^{-iEt}e^{i(k+\frac{1}{2})\theta}}{E+|V|+m}\nonumber\\
\big{[}\frac{1-2k}{s}(\alpha J_{k-\frac{1}{2}}(s)+\beta Y_{k-\frac{1}{2}}(s))+\nonumber\\
(\alpha J_{k-\frac{3}{2}}(s)+\beta Y_{k-\frac{3}{2}}(s))\big{]}~.
\end{eqnarray}
\paragraph{$-\kappa^2=(E+|V|)^2-m^2<0$.} 
In that case, the equation is
\be
r^2\partial^2\psi'_1-r^2\kappa^2\psi'_1+k(1-k)\psi'_1=0~.
\ee
If we introduce again $s=r\kappa$ and $\psi'_1=\sqrt{s}\Phi_1$, we find the following equation:
\be
s^2\partial^2_s\Phi_1+s\partial_s\Phi_1-s^2\Phi_1-n^2\Phi_1=0~.
\ee
It admits modified Bessel functions of the first and second kind as solutions (denoted by $I_{k-\frac{1}{2}}$ and $K_{k-\frac{1}{2}}$).

We find that
\be\label{ext1}
\psi_1=\sqrt{\kappa}e^{-iEt}e^{i(k-\frac{1}{2})\theta}(\alpha' I_{k-\frac{1}{2}}(s)+\beta' K_{k-\frac{1}{2}}(s))
\ee
and that 
\begin{eqnarray}\label{ext2}
\psi_2=-i\kappa^{\frac{3}{2}}\frac{e^{-iEt}e^{i(k+\frac{1}{2})\theta}}{E+|V|+m}\nonumber\\
\big{[}\frac{1-2k}{s}(\alpha' I_{k-\frac{1}{2}}(s)+\beta' K_{k-\frac{1}{2}}(s))+\nonumber\\
(\alpha' I_{k-\frac{3}{2}}(s)-\beta' K_{k-\frac{3}{2}}(s))\big{]}~.
\end{eqnarray}

The next step is to choose some numerical values for the potential $V_0$ and the radius of the spherical box 
and find the energy eigenvalues numerically. In order to do that, we impose that the ratio 
$\frac{\psi_1}{\psi_2}$ coincides at the boundary $r=R'$.
\subsection{Simulations}
We choose $m=1$, $R'=5$ and $|V_0|=1$. 
We are going to superpose energy eigenstates whose eigenvalues $E$ satisfy the relations
\be\label{domain}
E^2\leq m^2\textrm{ and }(E+|V_0|)^2-m^2\geq 0~.
\ee 
Therefore we take
\begin{itemize}
\item Eq. (\ref{int1}) and Eq. (\ref{int2}) for the interior solution (and we set $\beta=0$ for the solutions to be regular at the origin)~,
\item Eq. (\ref{ext1}) and Eq. (\ref{ext2}) for the exterior solution (and we set $\alpha'=0$ for the solutions to vanish at $r=\infty$.)
\end{itemize}
The next step is to find the energy eigenvalues numerically, for different values of $k$, by imposing that
\be
\frac{\psi_{{\rm in},1}}{\psi_{{\rm in},2}}\bigg|_{r=5}=\frac{\psi_{{\rm out},1}}{\psi_{{\rm out},2}}\bigg|_{r=5}
\ee
and by looking in the domain defined by Eq. (\ref{domain}). The last step is to find $\beta'$ by requiring that
\be
\psi_{{\rm in},1}\big|_{r=5}=\psi_{{\rm out},1}\big|_{r=5}~.
\ee
The guiding spinor used for the simulation is a superposition of 6 positive-energy eigenstates with different values of $k$. 
The details about the spinor can be found in Appendix \ref{app1}.

\begin{figure}  
\includegraphics[width=0.5\textwidth]{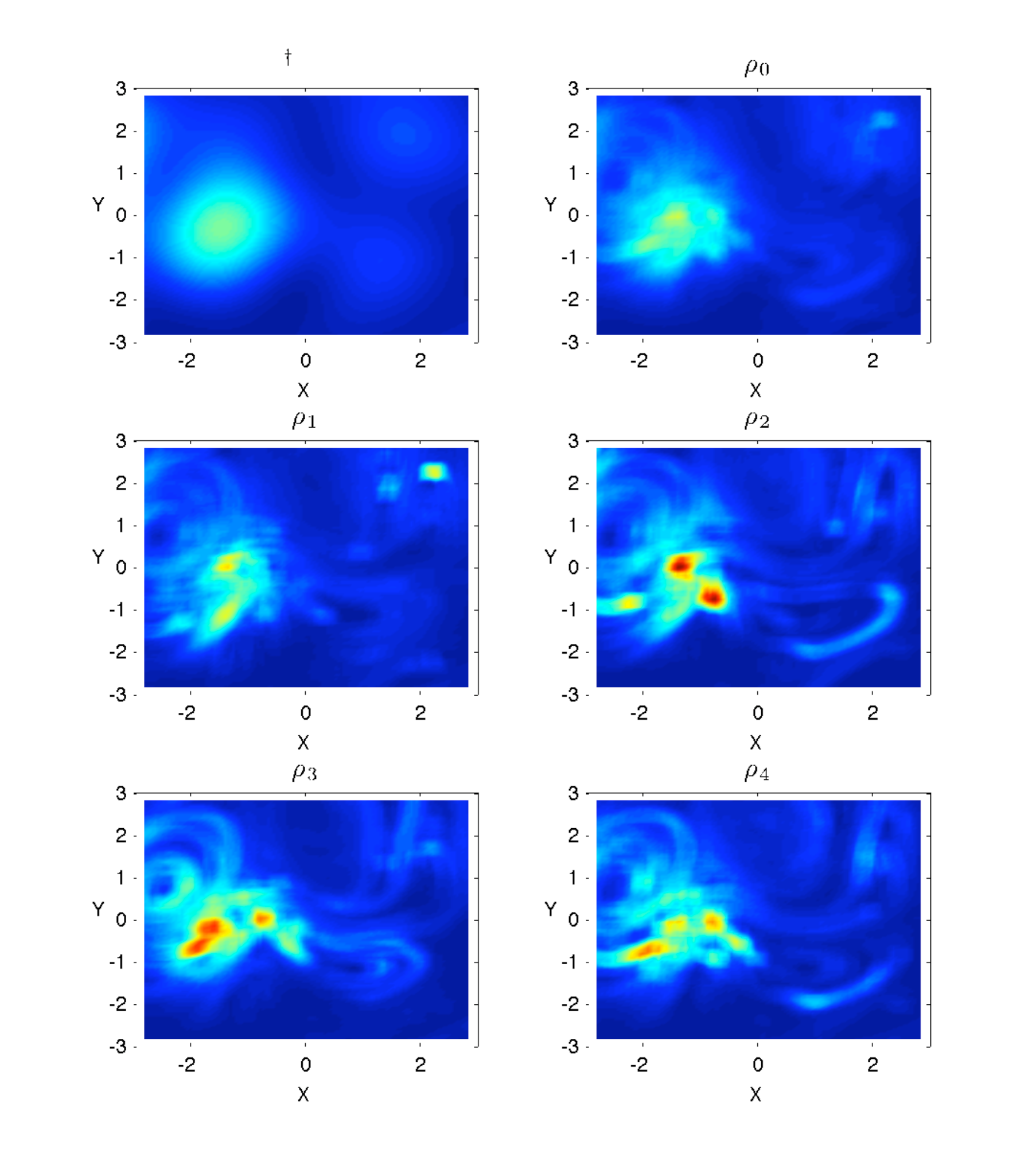}
\caption{\label{fig6a}Density plots. The first plot displays the smoothly coarse-grained standard density ($\widetilde{\psi^\dagger\psi}$) at $t=100$, 
while the 5 remaining plots show the smoothly coarse-grained non-equilibrium densities at $t=100$.
The initial non-equilibrium distributions ($\widetilde{\rho}_j$, with $j\in\{0,1,2,3,4\}$) are defined 
at Eq. (\ref{rho0}) and Eq. (\ref{rhoj}). 
The guiding spinor $\psi$ is the superposition of $6$ modes given in Appendix \ref{app1}.}
\end{figure}

\begin{figure}
\includegraphics[width=0.5\textwidth]{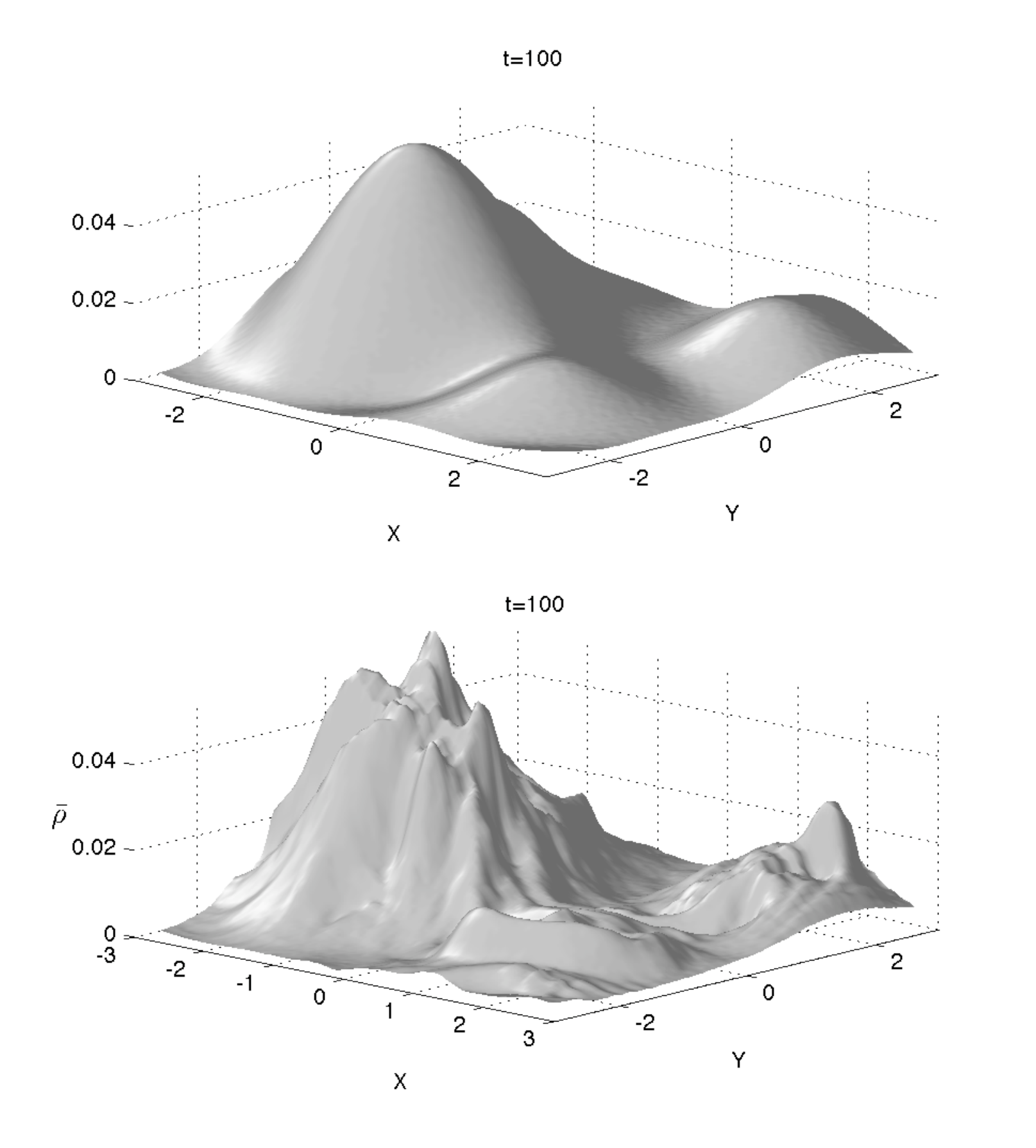}
\caption{\label{fig6b}Surface plot for $\widetilde{\psi^\dagger\psi}(t=100)$ and $\widetilde{\rho_0}(t=100)$ (\mbox{$\widetilde{\rho_0}(t=0)$} being defined at Eq. (\ref{rho0})).
The guiding spinor $\psi$ is the superposition of $6$ energy eigenstates given in Appendix \ref{app1}.}
\end{figure}

For the relaxation simulation, we have considered the evolution of the 5 previous non-equilibrium distributions for $t=50$ and $t=100$. 
The lattice consists of $1024\times 1024$ points distributed uniformly in the box $[-3,3]\times[3,3]$ and the precision on backtracked trajectories is 
$0.005$. If the box in divided in $32\times32$ CG cells, the mean percentage 
of good trajectories at $t=100$ is $93.22\%$ and the worst cell has a percentage of $67.58\%$.  Some of these results are illustrated in Fig. (\ref{fig6a}) and 
Fig. (\ref{fig6b}).
\subsection{Discussion}
While the relaxation to quantum equilibrium is slower in this case (compared to the Dirac oscillator case), 
it must be pointed out that the coarse-graining length is smaller here ($3/5$ of the coarse-graining length used in the previous case). 
Also, if the relaxation time depends on the energy spreading, we would expect a longer relaxation time in this case anyway. 
Finally, the Dirac oscillator has a strong spin-orbit coupling \cite{diracosc3d1} which may speed up the relaxation.

\begin{figure}
\includegraphics[width=0.5\textwidth]{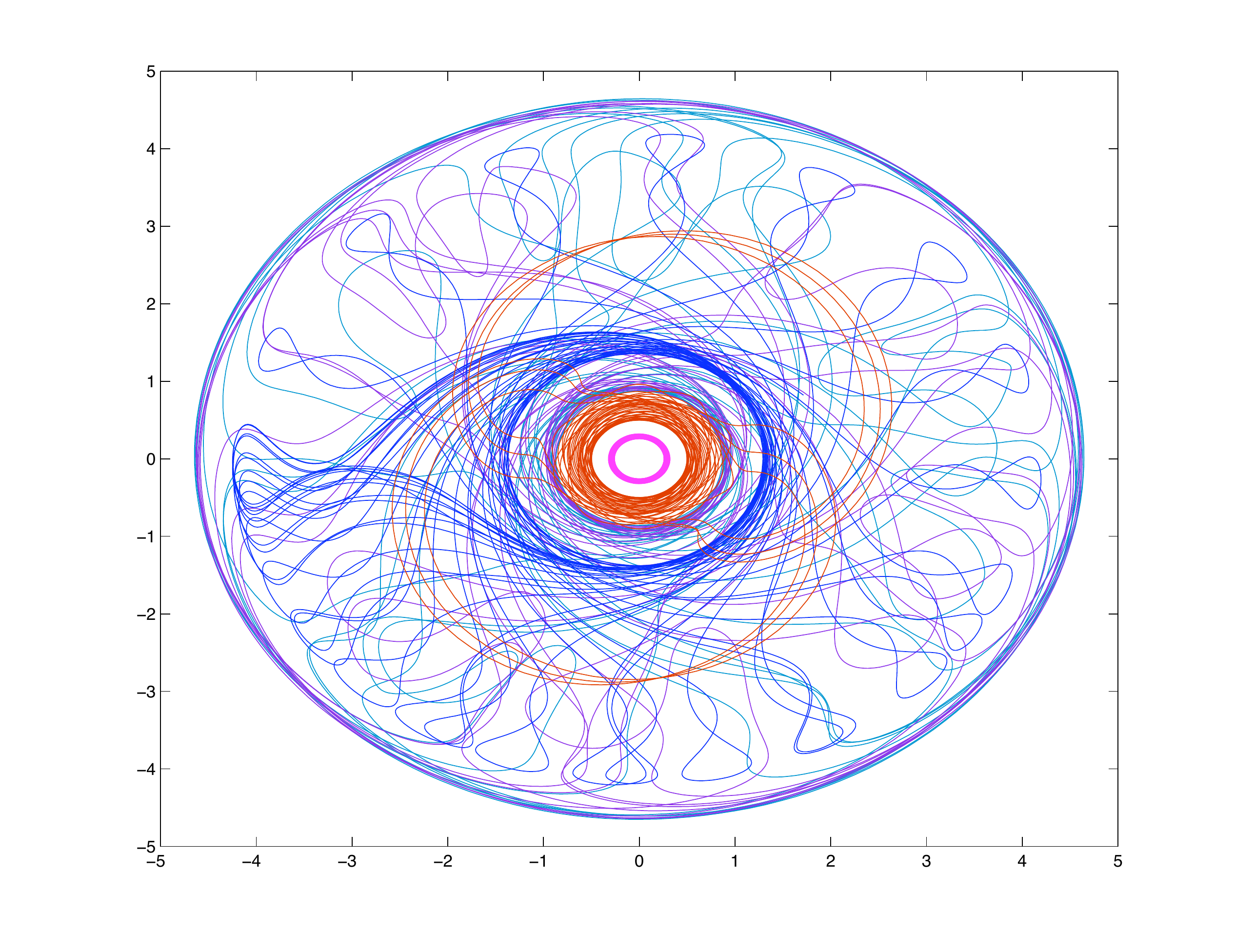}
\caption{\label{fig7}Various trajectories originating from different final positions and 
backtracked from $t=1000$ to $t=0$. 
The guiding spinor is the superposition of $3$ energy eigenstates with positive values of $k$ given in Appendix \ref{app1}.}
\end{figure}

Another interesting example is the case of a wave-function with only positive values of $k$ (which means that the 
direction of rotation doesn't change). In that case, non-equilibrium distributions with support around the origin will not relax. In order to understand 
this, we plot a few backtracked trajectories (cf Fig. \ref{fig7}) for a superposition of the $3$ modes with positive values of $k$ .
As we can see, trajectories from the inner core are backtracked to the inner core, which is a bad thing for relaxation. 
One might argue that only the modes with $J_0(x)$ are dominant in that region (these are the eigenstates with $k=1/2$ and $k=3/2$). 
We have only tested that the inner core is backtracked to the inner core for a superposition of $3$ modes with positive values of $k$ 
but it might be a general feature of many superpositions with positive values of $k$. 
For instance, if we simply add one mode with $k=-1/2$, the core is not disconnected anymore: trajectories backtracked from the core can end up in the outer region.
\section{Conclusion}
We have considered two 2-dimensional systems (the Dirac oscillator and the Dirac particle in a spherical box) and we have done 
numerical simulations which show an efficient relaxation to quantum equilibrium for Dirac fermions, 
provided that enough modes are superposed in order to get sufficient complexity in the dynamics. 
The relaxation takes place despite the absence of nodes thus it can be attributed to the intrinsic vorticity of the de Broglie-Bohm Dirac velocity field.
For the second system, we have also given an example of a non-trivial spinor (with positive values of $k$) for which there will never be complete relaxation.

This last example does not ruin the idea of relaxation. 
Firstly it is a two-dimensional system.
But secondly (and more importantly), even in a two-dimensional universe, this system does not sit there on its own: it starts to interact 
with other systems, resulting in a more complex system, where relaxation can take place.
However this example is interesting with respect to relic non-equilibrium particles \cite{valentini07}, which are hypothetical non-equilibrium 
particles from the very early universe that would still be in non-equilibrium today. The possibility of finding such systems arises 
because the spatial expansion in the very early universe 
stretches non-equilibrium length-scales and competes with the natural process of relaxation. Then, if these non-equilibrium particles 
decouple very early and if they are still in non-equilibrium at the time of decoupling, 
they can't be driven to quantum equilibrium by interaction with other systems. Furthermore, if they are in an energy eigenstate, the natural 
relaxation does not take place and the non-equilibrium can be preserved up to the present day. The last example shows 
that non-equilibrium could also be preserved when the spinor is more complex than an energy eigenstate.

As possible future work, one can study
\begin{itemize}
\item the influence of spatial expansion (which is relevant for the early universe and results in a stretching of the non-equilibrium length-scales), 
\item the role of mass, since it is not fundamental and beables should really be attributed to massless fermions (discussed in \cite{cowi}), 
\item relaxation timescales, via the H-function (like in \cite{toruva}), 
\item and the generalization to $3+1$-dimensional systems, in particular spinning systems for which the rotation direction does not change.
\end{itemize}
Also the Dirac equation is not the only relativistic equation. For instance, it is still an open question whether neutrinos are Dirac or Majorana fermions  
(a Majorana particle being its own anti-particle). The Majorana equation is not only relevant for particle physics but also to quantum information and condensed matter research 
(an overview of the wide relevance of the Majorana equation is given in \cite{wilczek}). From the point of view of the pilot-wave theory, the Majorana equation is also very particular.
Indeed, in a forthcoming paper \cite{colin-majorana}, we will show that the Majorana equation predicts luminal motion for the beable, although the Majorana 
solutions involve the mass. In the same paper, we study relaxation to quantum equilibrium for systems governed by the Majorana equation.

\begin{acknowledgments}
I thank Howard Wiseman for comments and suggestions on a previous manuscript.
I acknowledge financial support from a Perimeter Institute Australia Foundations postdoctoral research fellowship. 
This work was also supported by the Australian Research Council Discovery Project DP0880657, ``Decoherence, Time-Asymmetry and the Bohmian View on the 
Quantum World''.
Research at Perimeter Institute is funded by the Government of Canada through Industry Canada and by the Province of Ontario through the Ministry of Research and Innovation.
The relaxation simulation for the Dirac oscillator was performed on the \textit{Griffith University HPC Cluster - V20z}.
\end{acknowledgments}

\appendix
\section{Guiding spinors for the second system}\label{app1}
We take $|V_0|=1$, $m=1$ and $R'=5$.

For an energy eigenstate, the interior solution is 
\begin{eqnarray}
\psi^{in}_1(r,\theta)=\kappa_{in}^{\frac{1}{2}}e^{-iEt}e^{i(k-\frac{1}{2})\theta} J_{k-\frac{1}{2}}(\kappa_{in} r)\\
\psi^{in}_2(r,\theta)=-i\kappa_{in}^{\frac{3}{2}}\frac{e^{-iEt}e^{i(k+\frac{1}{2})\theta}}{E+|V_0|+m}\nonumber\\
\big{[}\frac{1-2k}{\kappa_{in} r}J_{k-\frac{1}{2}}(\kappa_{in} r)+J_{k-\frac{3}{2}}(\kappa_{in} r)\big{]}~,
\end{eqnarray}
where $\kappa_{in}=\sqrt{(E+|V_0|)^2-m^2}$, while the exterior solution ($r\geq R'$) is 
\begin{eqnarray}
\psi^{out}_1(r,\theta)=\kappa_{out}^{\frac{1}{2}}e^{-iEt}e^{i(k-\frac{1}{2})\theta}\beta' K_{k-\frac{1}{2}}(\kappa_{out} r)\\
\psi^{out}_2(r,\theta)=-i\kappa_{out}^{\frac{3}{2}}\frac{e^{-iEt}e^{i(k+\frac{1}{2})\theta}}{E+m}\nonumber\\
\big{[}\frac{1-2k}{\kappa_{out} r}\beta' K_{k-\frac{1}{2}}(\kappa_{out} r)-\beta' K_{k-\frac{3}{2}}(\kappa_{out} r)\big{]}~
\end{eqnarray}
with $\kappa_{out}=\sqrt{m^2-E^2}$.

The energy eigenvalues are found by matching the ratio $\frac{\psi_1}{\psi_2}$ for the interior and exterior solutions:
\begin{eqnarray}
\begin{tabular}{| l | l |}
\hline
$k$ & $E$\\
\hline
$1/2$ & $0.410077354998218$\\
$3/2$ & $0.610542082182398$ \\
$5/2$ & $0.812057491976715$\\
$-1/2$ & $0.598385922365134$ \\ 
$-3/2$ & $0.356509811273382$\\
$-5/2$ & $0.510184308650916$ \\
\hline
\end{tabular}~.
\end{eqnarray}

The parameter $\beta'$ are found by matching $\psi_1$ for the interior and exterior solutions:
\begin{eqnarray}
\begin{tabular}{| l | l |}
\hline
$k$ & $\beta'$\\
\hline
$1/2$ & $-32.6316901377613$\\
$3/2$ & $-19.79344405979468$ \\
$5/2$ & $-5.13915809445641$\\
$-1/2$ & $22.59183163168054$ \\ 
$-3/2$ & $24.1846971959765$\\
$-5/2$ & $-12.1855792791713$ \\
\hline
\end{tabular}~.
\end{eqnarray}
\\
Finally we take random phases $e^{i\phi}$:
\begin{eqnarray}
\begin{tabular}{| l | l |}
\hline
$k$ & $\phi$\\
\hline
$1/2$ & $0.797881698340871$\\
$3/2$ & $5.73890975922526$ \\
$5/2$ & $3.97323032474265$\\
$-1/2$ & $1.74985591686112$ \\ 
$-3/2$ & $5.11905989575681$\\
$-5/2$ & $0.61286443954863$ \\
\hline
\end{tabular}~.
\end{eqnarray}
\\
The superposition of $6$ modes is obtained by summing the corresponding eigenstates multiplied by the phase coefficients. In the end the spinor is normalized 
numerically. The superposition of $3$ (resp. $4$) modes is obtained by superposing only the first $3$ (resp. $4$) eigenstates.

\begin{thebibliography}{10}

\bibitem{debroglie28}
{L.\ de Broglie, in ``Electrons et Photons: Rapports et Discussions du
  Cinqui\`eme Conseil de Physique'', Gauthier-Villars, Paris, 105 (1928),
  English translation: G.\ Bacciagaluppi and A.\ Valentini, ``Quantum Theory at
  the Crossroads: Reconsidering the 1927 Solvay Conference'', Cambridge
  University Press (2009), also quant-ph/0609184.}

\bibitem{bohm521}
{D.\ Bohm, {\em Phys.\ Rev.}\ {\bf 85}, 166 (1952).}

\bibitem{bohm522}
{D.\ Bohm, {\em Phys.\ Rev.}\ {\bf 85}, 180 (1952).}

\bibitem{costva}
S.~Colin, W.~Struyve, and A.~Valentini.
\newblock {Instability of Bohm's dynamics}.
\newblock 2011.

\bibitem{valentini91a}
{A.\ Valentini, {\em Phys.\ Lett.\ A} {\bf 156}, 5 (1991).}

\bibitem{valentini91b}
{A.\ Valentini, {\em Phys.\ Lett.\ A} {\bf 158}, 1 (1991).}

\bibitem{durr92}
{D.\ D\"urr, S.\ Goldstein and N.\ Zangh\`\i, {\em J.\ Stat.\ Phys.}\ {\bf 67},
  843 (1992) and quant-ph/0308039.}

\bibitem{valentini-phd}
{A.\ Valentini, ``On the Pilot-Wave Theory of Classical, Quantum and Subquantum
  Physics", PhD.\ Thesis, International School for Advanced Studies, Trieste
  (1992), online \url{http://www.sissa.it/ap/PhD/Theses/valentini.pdf}.}

\bibitem{bricmont}
{J.\ Bricmont, in {\em Chance in Physics: Foundations and Perspectives}, eds.\
  J.\ Bricmont, D.\ D\"urr, M.C.\ Galavotti, G.\ Ghirardi, F.\ Petruccione and
  N.\ Zangh\`i, Lecture Notes in Physics 574, Springer-Verlag, Berlin, 39
  (2001).}

\bibitem{wiseman2007}
{H. M. Wiseman, {\em New J. Phys.} {\bf 9}, 165 (2007).}

\bibitem{valentini08}
Antony Valentini.
\newblock Inflationary cosmology as a probe of primordial quantum mechanics.
\newblock {\em Phys. Rev. D}, 82(6):063513, Sep 2010.

\bibitem{cowi}
S.~Colin and H.~M. Wiseman.
\newblock The zig-zag road to reality.
\newblock {\em J. Phys. A: Math. Theor.}, 44(34):345304, Aug 2011.

\bibitem{valentini05}
{A.\ Valentini and H.\ Westman, {\em Proc.\ R.\ Soc.\ A} {\bf 461}, 253 (2005)
  and quant-ph/0403034.}

\bibitem{cost10}
{S.\ Colin and W.\ Struyve, {\em New\ J.\ Phys.} {\bf 12}, 043008 (2010).}

\bibitem{toruva}
{Towler, M. D. and Russell, N. J. and Valentini, A., Timescales for dynamical
  relaxation to the Born rule, arXiv:1103.1589}.~Accepted for publication~in
  {\em Proc. Roy. Soc. A}.

\bibitem{frisk97}
{H.\ Frisk, {\em Phys.\ Lett.\ A} {\bf 227}, 139 (1997).}

\bibitem{efthymiopoulos}
C.~Efthymiopoulos, C.~Kalapotharakos, and G.~Contopoulos.
\newblock Origin of chaos near critical points of quantum flow.
\newblock {\em Phys. Rev. E}, 79(3):036203, Mar 2009.

\bibitem{deotto98}
{E.\ Deotto and G.C.\ Ghirardi, {\em Found.\ Phys.}\ {\bf 28}, 1 (1998) and
  quant-ph/9704021.}

\bibitem{bohm53}
{D.\ Bohm, {\em Prog.\ Theor.\ Phys.}\ {\bf 9}, 273 (1953).}

\bibitem{bohm93}
{D.\ Bohm and B.J.\ Hiley, {\em The Undivided Universe}, Routledge, New York
  (1993).}

\bibitem{moshinsky}
M.~Moshinsky and Szczepaniak A.
\newblock {The Dirac oscillator}.
\newblock {\em J. Phys. A: Math. Gen.}, 22:L817--L819, June 1989.

\bibitem{diracosc3d1}
V.~I. Kukulin, G.~Loyola, and M.~Moshinsky.
\newblock {A Dirac equation with an oscillator potential and spin-orbit
  coupling}.
\newblock {\em Phys. Lett. A}, 58:19--22, May 1991.

\bibitem{diracosc3d2}
O.~Casta\~nos, A.~Frank, R.~L\'opez, and L.~F. Urrutia.
\newblock {Soluble extensions of the Dirac oscillator with exact and broken
  supersymmetry}.
\newblock {\em Phys. Rev. D}, 43(2):544--547, Jan 1991.

\bibitem{villalba}
Victor~M. Villalba.
\newblock {Exact solution of the two-dimensional Dirac oscillator}.
\newblock {\em Phys. Rev. A}, 49(1):586--587, June 1994.

\bibitem{press92}
{W.H.\ Press, S.A.\ Teukolsky, W.T.\ Vetterling and B.P.\ Flannery, {\em
  Numerical recipes in FORTRAN}, Cambridge University press, Cambridge (1992).}

\bibitem{greiner}
W.\ Greiner.
\newblock {\em Relativistic Quantum Mechanics: Wave Equations}.
\newblock Springer-Verlag, 1990.

\bibitem{valentini07}
{A.\ Valentini, {\em J.\ Phys.\ A} {\bf 40}, 3285 (2007) and hep-th/0610032.}

\bibitem{wilczek}
F.~Wilczek.
\newblock Majorana returns.
\newblock {\em Nature Physics}, 5:614--618, 2009.

\bibitem{colin-majorana}
{S. Colin, Pilot-wave theory for the Majorana equation.}

\end{thebibliography}
\end{document}